\documentclass[submission,copyright,creativecommons]{eptcs}
\usepackage{mathtools}
\usepackage{setspace}
\usepackage{ebproof}
\usepackage{stmaryrd}
\usepackage{amssymb}
\usepackage{tabularx}
\usepackage{marvosym}
\usepackage{breakcites}
\usepackage{hyperref}
\usepackage[normalem]{ulem}
\usepackage{listings}
\usepackage{xcolor}
\usepackage{float}

\providecommand{\event}{PLACES25} % Name of the event you are submitting to

\usepackage{iftex}

\ifpdf
  \usepackage{underscore}         % Only needed if you use pdflatex.
  \usepackage[T1]{fontenc}        % Recommended with pdflatex
\else
  \usepackage{breakurl}           % Not needed if you use pdflatex only.
\fi

\title{Thread and Memory-Safe Programming with CLASS}
\author{Lu\' is Caires
\institute{Instituto Superior T\' ecnico (U Lisboa) / INESC-ID}
\email{luis.caires@tecnico.ulisboa.pt}
}
\def\titlerunning{Thread and Memory-Safe Programming with CLASS}
\def\authorrunning{Lu\' is Caires}
\begin{document}
%%%%%%%%%%%%%%%%%%%%%%% %%%%%%%%%%%%%%%%
%%%%%%%%%%%%%%%%%%%%%%% %%%%%%%%%%%%%%%%
%%%%%%%%%%%% MY MACROS BEGIN %%%%%%%%%%%%%%%%
%%%%%%%%%%%%%%%%%%%%%%% %%%%%%%%%%%%%%%%
%%%%%%%%%%%%%%%%%%%%%%% %%%%%%%%%%%%%%%%
%
%
%
%
%Cell Modalities 
%\newcommand{\dboxbang}{\makebox[0pt][l]{{\vspace{-2em}$\dbox{}$}}\raisebox{.23ex}{\hspace{0.28em}{\scriptsize $!$}}\;}
%\newcommand{\dboxwhy}{\makebox[0pt][l]{$\diamondsuit$}\raisebox{.23ex}{\hspace{0.20em}{\scriptsize $?$}}\;}
%
\definecolor{pcolor}{RGB}{156,90,107}
\definecolor{tcolor}{RGB}{0,102,204}
\definecolor{turquoise}{RGB}{0,153,153}
\definecolor{light-blue}{RGB}{51,153,255}
\definecolor{dark-blue}{RGB}{0,0,204}
\definecolor{purple}{RGB}{76,0,153}
\definecolor{red}{RGB}{212,1,1}
\newcommand{\hide}[1]{}
\newif\ifprettysyntax
\prettysyntaxtrue
%
%
%!TEX encoding = UTF-8 Unicode
\ifprettysyntax
\newcommand{\zero}{{\textcolor{pcolor}{\mathsf 0}}}
\newcommand{\fwd}[2]{\textcolor{pcolor}{\mathsf{fwd} }\ #1 \ #2} 
\newcommand{\fwdA}[3]{\textcolor{pcolor}{\mathsf{fwd}}_{#1} \ #2 \ #3} 
\newcommand{\mix}[2]{#1 \ \textcolor{pcolor}{\mathsf{||}} \ #2}
\newcommand{\shareP}{\textcolor{pcolor}{\mathsf{share}}}
\newcommand{\cutP}{\textcolor{pcolor}{\mathsf{cut}}}
\newcommand{\mixP}{\textcolor{pcolor}{\mathsf{par}}}
\newcommand{\mixD}{\textcolor{pcolor}{\mathsf{||}}}
\newcommand{\cutD}[1]{\textcolor{pcolor}{\mathsf{|}} #1 \textcolor{pcolor}{\mathsf{|}}}
\newcommand{\cut}[3]{\cutP\ \{ #2   \ \textcolor{pcolor}{\mathsf{|}} #1 \textcolor{pcolor}{\mathsf{|}}   \ #3\} }
\newcommand{\cutB}[4]{#2.#3 \  \textcolor{pcolor}{\mathsf{|}}!#1\textcolor{pcolor}{\mathsf{|}} \ #4}
\newcommand{\share}[3]{\textcolor{pcolor}{\mathsf{share}} \ #1 \ \{  #2 \ \textcolor{pcolor}{||} \ #3 \}}
\newcommand{\mut}[3]{\textcolor{pcolor}{\mathsf{mut}} \ #1 \ \{  #2 \ \textcolor{pcolor}{||} \ #3 \}}
\newcommand{\nsum}[2]{#1 \  \textcolor{pcolor}{+} \ #2}
\newcommand{\pone}[1]{\textcolor{pcolor}{\mathsf{close}} \ #1}
\newcommand{\pbot}[2]{\textcolor{pcolor}{\mathsf{wait}} \ #1;#2}
\newcommand{\pparl}[3]{\textcolor{pcolor}{\mathsf{recv}} \ #1(#2);#3}
\newcommand{\potimess}[3]{\textcolor{pcolor}{\mathsf{send}} \ #1(#2);#3}  
\newcommand{\potimes}[4]{\textcolor{pcolor}{\mathsf{send}} \ #1(#2.#3);#4}  
\newcommand{\fout}[3]{\textcolor{pcolor}{\mathsf{send}} \ #1(#2);#3}
\newcommand{\pcase}[3]{\textcolor{pcolor}{\mathsf{case}} \ #1 \  \{ |\mathsf{inl} : #2, \ | \mathsf{inr}: #3\}} 
\newcommand{\gcase}[2]{\textcolor{pcolor}{\mathsf{case}} \ #1 \  \{ #2\}} 
\newcommand{\cleft}[2]{#1.\textcolor{pcolor}{\mathsf{inl}};#2}
\newcommand{\cright}[2]{#1.\textcolor{pcolor}{\mathsf{inr}};#2}
\newcommand{\choice}[3]{\mathsf{#1} \ #2; #3}
\newcommand{\pbang}[3]{\textcolor{pcolor}{\mathsf{!}}#1(#2);#3} 
\newcommand{\pwhy}[2]{\textcolor{pcolor}{\mathsf{?}}#1; #2}
\newcommand{\pcopy}[3]{\textcolor{pcolor}{\mathsf{call}} \ #1(#2); #3} 
\newcommand{\ncell}[3]{\textcolor{pcolor}{\mathsf{cell}} \ #1(#2.#3)}
\newcommand{\nfree}[1]{\textcolor{pcolor}{\mathsf{release}} \ #1} 
\newcommand{\rd}[3]{\textcolor{pcolor}{\mathsf{read}} \ #1(#2);#3}
\newcommand{\nwrt}[4]{\textcolor{pcolor}{\mathsf{wrt}} \ #1(#2.#3);#4} 
\newcommand{\fwrt}[3]{\textcolor{pcolor}{\mathsf{wrt}} \ #1 \ #2;#3}
\newcommand{\fwdE}[2]{\textcolor{pcolor}{\mathsf{fwd}^!} \ #1 \ #2}
\newcommand{\fwdEA}[3]{\textcolor{pcolor}{\mathsf{fwd}^!}_{#1} \ #2 \ #3}
\newcommand{\cellE}[3]{\textcolor{pcolor}{\mathsf{cell}_!} \ #1 \ #2 \ #3 }
\newcommand{\pexists}[3]{\textcolor{pcolor}{\mathsf{sendty}} \ #1(#2); #3}
\newcommand{\pforall}[3]{\textcolor{pcolor}{\mathsf{recvty}} \ #1(#2);#3} 
\newcommand{\lock}[2]{\textcolor{pcolor}{\mathsf{lock}} \ #1;#2} 
\newcommand{\unlock}[2]{\textcolor{pcolor}{\mathsf{unlock}} \ #1;#2} 
\newcommand{\shareL}[3]{\textcolor{pcolor}{\mathsf{shareL}} \ #1 \ \{  #2 \ \textcolor{pcolor}{||} \ #3  \}}
\newcommand{\shareR}[3]{\textcolor{pcolor}{\mathsf{shareR}} \ #1 \ \{   #2 \ \textcolor{pcolor}{||} \ #3  \}}
\newcommand{\mlet}[2]{\textcolor{pcolor}{\mathsf{let}} \ #1 \ #2}
\newcommand{\mletB}[2]{\textcolor{pcolor}{\mathsf{let!}} \ #1 \ #2}
\newcommand{\mif}[3]{\textcolor{pcolor}{\mathsf{if}} \ (#1)\{#2\}\{#3\}}
\else 
\newcommand{\zero}{{\mathbf 0}}
\newcommand{\fwd}[2]{#1 \leftrightarrow #2} 
\newcommand{\mix}[2]{#1  \parallel   #2}
\newcommand{\cut}[3]{#2  \parallel_{#1}  #3}
\newcommand{\cutB}[4]{#2.#3   \parallel_{!#1} #4}
\newcommand{\share}[3]{{#1.\mathsf{share}}[#2\;|\;#3]}
\newcommand{\nsum}[2]{#1  + #2}
\newcommand{\pone}[1]{\dual{#1}.\mathsf{close}}
\newcommand{\pbot}[2]{#1.\mathsf{close};#2} 
\newcommand{\pparl}[3]{#1(#2){;}#3}
\newcommand{\potimes}[4]{\dual{#1}(#2.#3);#4}
\newcommand{\fout}[3]{\dual{#1}\langle #2 \rangle{;} #3}
\newcommand{\pcase}[3]{#1.\mathsf{case}[#2, #3]} 
\newcommand{\cleft}[2]{\dual{#1}.\mathsf{inl};#2}
\newcommand{\cright}[2]{\dual{#1}.\mathsf{inr};#2} 
\newcommand{\pbang}[3]{!#1(#2.#3)} 
\newcommand{\pwhy}[2]{?#1;#2}
\newcommand{\pcopy}[3]{\dual{#1}?(#2);#3} 
\newcommand{\ncell}[3]{#1.\mathsf{cell}(#2. #3)}
\newcommand{\rd}[3]{#1.\mathsf{rd}(#2){;}#3}
\newcommand{\nwrt}[4]{#1.\mathsf{wr}(#2.#3){;}#4} 
\newcommand{\fwrt}[3]{#1.\mathsf{wr}\langle #2 \rangle; #3}
\newcommand{\fwdE}[2]{#1 \leftrightarrow_! #2}
\newcommand{\cellE}[3]{#1.\mathsf{cell}(\fwdE{#2}{#3})}
\newcommand{\pexists}[3]{\dual{#1}\langle #2 \rangle; #3}
\newcommand{\pforall}[3]{#1(#2);#3} 
\newcommand{\lock}[2]{#1.\mathsf{lock};#2} 
\newcommand{\unlock}[2]{#1.\mathsf{unlk};#2} 
\newcommand{\shareL}[3]{#1.\mathsf{shareL}[#2 \;|\;  #3]} 
\newcommand{\shareR}[3]{#1.\mathsf{shareR}[#2 \;|\;  #3]} 
\fi
\newcommand{\deff}{\triangleq}
\newcommand{\LC}[1]{{\color{red}#1}}
\newcommand{\PR}[1]{{\color{blue}#1}}
\newcommand{\tto}[0]{\Rightarrow}
\newcommand{\sep}[0]{\;|\;}
\newcommand{\dual}[1]{\overline{#1}}
\newcommand{\subs}[3]{\{#1/#2\}#3} 
\newcommand{\texists}[2]{\exists #1. #2}
\newcommand{\tforall}[2]{\forall #1.#2} 
\newcommand{\with}{\mathbin{\binampersand}}
%Locks 
\colorlet{dgreen}{green!40!black}
\newcommand{\cboxwhy}[1]{{\textcolor{dgreen}{\boxwhy}} #1} 
\newcommand{\clboxwhy}[1]{{\textcolor{red}{\lboxwhy}} #1} 
\newcommand{\cboxbang}[1]{{\textcolor{dgreen}{\boxbang}} #1}
\newcommand{\clboxbang}[1]{{\textcolor{red}{\lboxbang}} #1}
\newcommand{\lboxbang}{\makebox[0pt][l]{$\boxslash$}\raisebox{0ex}{\hspace{0.27em}{\scriptsize $!$}}\;}
\newcommand{\lboxwhy} {\makebox[0pt][l]{$\boxslash$}\raisebox{0ex}{\hspace{0.21em}{\scriptsize $?$}}\;}
\newcommand{\boxbang}{\makebox[0pt][l]{$\boxempty$}\raisebox{0ex}{\hspace{0.27em}{\scriptsize $!$}}\;}
\newcommand{\boxwhy} {\makebox[0pt][l]{$\boxempty$}\raisebox{0ex}{\hspace{0.21em}{\scriptsize $?$}}\;}
\newcommand{\inter}[3]{ \mathcal I_{#1}(#2, #3)}
\newcommand{\ptypes}[0]{PaT}
%
%%%listing settings begin 
\lstset{
basicstyle=\linespread{1.0}\ttfamily\mdseries
\small,
columns=fullflexible,
    literate=*{?}{{\textcolor{pcolor}{?}}}{1}
    		 {*}{{\textcolor{pcolor}{*}}}{1}
    		 {+}{{\textcolor{pcolor}{+}}}{1}
    		 {<-}{{\textcolor{pcolor}{<- }}}{1}
    		 {->}{{\textcolor{pcolor}{-> }}}{1}
    		 {!}{{\textcolor{pcolor}{!}}}{1}
    		 {\#}{{\textcolor{pcolor}{\#}}}{1}
    		 {[]}{{\textcolor{pcolor}{()}}}{1}
%    		 {~}{{\textcolor{pcolor}{${\mathsf{\sim}}$}}}{1},
    		{~}{{\fontfamily{ptm}\selectfont\textcolor{pcolor}{{\textasciitilde}} }}1,
  }
\lstdefinelanguage{CLASS} {
identifierstyle=\color{blue},
morekeywords ={[2]proc,type,rec,gen_rec, and}, sensitive = true, 
morekeywords={[3]close, wait, case, pair, of, offer, choice, recv, send, state, statel, usage, usagel, affine, coaffine, call,  letc, drop, use, discard,
take, put, release, lint, colint, let, print, println, if, then, else,
sendty, recvty, cell, usage, free, rd, wrt, lock, unlk, par, cut, fwd, share}, sensitive=false,
morekeywords={[4]\~{}}, sensitive=true,
morekeywords={[4]()}, sensitive=true,
keywordstyle={[1]\color{pcolor}},
keywordstyle={[2]\color{red}},
keywordstyle={[3]\color{pcolor}},
keywordstyle={[4]\color{pcolor}},
morecomment=[l]{--},
morecomment=[s]{/*}{*/},
morestring=[b]", }
\lstset{language=CLASS}
%%%listing settings end s
%
%normal forms 
\newcommand{\keyc}[1]{{\textcolor{pcolor}{\texttt{#1}}}} 
\newcommand{\keyi}[1]{{\textcolor{blue}{\texttt{#1}}}} 
\newcommand{\keyt}[1]{{\textcolor{red}{\texttt{#1}}}} 

\newcommand{\nf}[1]{\llbracket #1 \rrbracket} 
\newcommand{\nfwhy}[1]{\llbracket #1 \rrbracket_?}
%has an atomic process ready to interact 
\newcommand{\obs}[2]{#1 \downarrow_{#2}}
\newcommand{\nshare}[2]{\text{nshare}(#1, #2)}
%%offers an action 
\newcommand{\offers}{\vartriangleright} 
%logical contexts 
\newcommand{\linctxt}[1]{\lpred{#1}}
\newcommand{\unrctxt}[1]{{\lpred{#1}^!}}
%true and false 
\newcommand{\true}{\text{true}}
\newcommand{\false}{\text{false}}
%liveness predicate 
%\newcommand{\live}[1]{\text{live}(#1)} 
%reduction strategy relations 
\newcommand{\cuteval}[1]{\lto{#1}} 
\newcommand{\ndexp}[1]{\lto{#1}_\epsilon}
\newcommand{\shexp}{\to_\epsilon}
%proof notations 
\newcommand{\dzero}{T\zero}
\newcommand{\dfwd}[2]{T\text{fwd} \ #1#2}
\newcommand{\dmix}[2]{T\text{mix} \ #1 #2}
\newcommand{\dcut}[3]{T\text{cut} \ (#1)#2#3}
\newcommand{\dshare}[3]{Tshare \ #1#2#3}
\newcommand{\dsum}[2]{T\text{sum} \ #1#2} 
\newcommand{\dcutB}[4]{Tcut! \ #1(#2)#3#4}
\newcommand{\dcopy}[3]{Tcopy \ #1(#2)#3}
\newcommand{\done}[1]{T\one \ #1}
\newcommand{\dbot}[2]{T\bot \ #1#2}
\newcommand{\dotimes}[4]{T\otimes \ #1(#2)#3#4}
\newcommand{\dparl}[3]{T\parl \ #1(#2)#3}
\newcommand{\dbang}[3]{T! \ #1(#2)#3}
\newcommand{\dwhy}[2]{T? \ #1#2}
\newcommand{\dcell}[3]{Tcell \ #1(#2)(#3)}
\newcommand{\dfree}[2]{Tfree \ #1#2} 
\newcommand{\drd}[3]{Tread \ #1(#2)#3} 
\newcommand{\dwrt}[4]{Twrt \ #1(#2)#3#4}
\newcommand{\doplusl}[2]{T\oplus_l \ #1#2}
\newcommand{\doplusr}[2]{T\oplus_r \ #1#2}
\newcommand{\dand}[3]{T\& \ #1#2#3}
\newcommand{\tot}{\xrightarrow[]{*}}
\newcommand{\toc}{\to_{\mathsf{c}}}
\newcommand{\toct}{\xrightarrow[]{*}_{\mathsf{c}}}
\newcommand{\equivc}{\equiv_{\mathsf{c}}}
\newcommand{\toS}{\mathrel{\mathrlap{\rightarrow}\mkern1mu\rightarrow}}
\newcommand{\toSt}{\mathrel{\mathrlap{\xrightarrow[]{*}}\mkern1mu\rightarrow}}
\newcommand{\fto}{\rightarrowdbl}
\newcommand{\ftoS}{\xrightarrowdbl[]{*}}
\newcommand{\lto}[1]{\xrightarrow[]{#1}}
\newcommand{\llto}[1]{\xRightarrow[]{#1}}
\newcommand{\ltos}[1]{\xrightarrow[s]{#1}}
\newcommand{\ltoS}[1]{\xRightarrow[]{#1}}
\newcommand{\lpred}[1]{\llbracket #1 \rrbracket} 
\newcommand{\lpreds}[1]{\lpred{#1}_s}
\newcommand{\lpredv}[1]{\lpred{#1}_{\to}}
\newcommand{\primeD}[1]{\mathcal P_r(#1)} 
\newcommand{\interE}[5]{\mathcal I^{\rightleftarrows}(#1,#2,#3,#4,#5)}
\newcommand{\rightarrowdbl}{\rightarrow\mathrel{\mkern-14mu}\rightarrow}
\newcommand{\xrightarrowdbl}[2][]{%
  \xrightarrow[#1]{#2}\mathrel{\mkern-14mu}\rightarrow
}
%Types 
\newcommand{\one}{{\mathbf1}}
\newcommand{\parl}{\mathbin{\bindnasrepma}}
%LTS
\newcommand{\aotimes}[3]{\dual{#1}(#2.#3)} 
\newcommand{\aclose}[1]{#1.\text{close}}
\newcommand{\ain}[2]{#1\langle #2 \rangle}
\newcommand{\arin}[2]{#1!\langle #2 \rangle}
\newcommand{\awhy}[1]{?#1}
\newcommand{\afree}[1]{#1.\text{free}}
\newcommand{\ard}[2]{#1.\text{read}\langle #2 \rangle}
\newcommand{\awrt}[2]{#1.\text{write}\langle #2 \rangle}
\newcommand{\aleft}[1]{#1.\text{inl}} 
\newcommand{\aright}[1]{#1.\text{inr}} 
%
%logical predicate stuff 
\newcommand{\predclose}{\mathcal{P}_\text{close}}
\newcommand{\predcom}[2]{\mathcal{P}_\text{com}(#1, #2)} 
\newcommand{\predcomb}[3]{\mathcal{P}_\text{com!}(#1, #2, #3)}
\newcommand{\predfree}{\mathcal{P}_\text{free}}
\newcommand{\predread}[3]{\mathcal{P}_\text{read}(#1, #2, #3)}
\newcommand{\predwrite}[3]{\mathcal{P}_\text{write}(#1, #2, #3)}
\newcommand{\enlto}[1]{\xrightarrow[+]{#1}} 

\newenvironment{centermath}
 {\begin{center}$\displaystyle}
 {$\end{center}}
 %Language names 
\newcommand{\muCLL}{$\mu\mathsf{CLL}$}
\newcommand{\piCLL}{$\mu\mathsf{CLL}$}
\newcommand{\LSS}{$\pi$SS} 
\newcommand{\LSSm}{$\pi$SS$_{\exists, \forall}$} 
\newcommand{\LLocks}{$\pi$SSL}
 %Linear Resources 
\newcommand{\emp}[1]{#1.\mathsf{emp}}
\newcommand{\acq}[3]{#1.\mathsf{acq}(#2);#3} 
\newcommand{\rel}[4]{#1.\mathsf{rel}(#2. #3);#4} 
\newcommand{\frel}[3]{#1.\mathsf{rel}\langle #2 \rangle;#3}
\newcommand{\res}[3]{#1.\mathsf{res}(#2.#3)} 
\newcommand{\fres}[2]{#1.\mathsf{res}\langle #2 \rangle} 
\newcommand{\nres}[5]{#1.\mathsf{res}(#2. #3) \mapsto #4.#5} 
\newcommand{\nemp}[3]{#1.\mathsf{empty} \mapsto #2.#3} 
%Linear Resources Modalities 
\newcommand{\cboxwhyf}[1]{{\textcolor{dgreen}{\boxwhy_f}} #1} 
\newcommand{\clboxwhyf}[1]{{\textcolor{red}{\lboxwhy_f}} #1} 
\newcommand{\cboxbangf}[1]{{\textcolor{dgreen}{\boxbang_f}} #1}
\newcommand{\clboxbangf}[1]{{\textcolor{red}{\lboxbang_f}} #1}
%List
\newcommand{\listT}[1]{\mathsf{LIST} \ #1} 
\newtheorem{innerlemma}{Lemma}
\newenvironment{mylemma}[1]
{\renewcommand\theinnerlemma{#1}\innerlemma}
{\endinnerlemma}
\newtheorem{innertheorem}{Theorem}
\newenvironment{mytheorem}[1]
{\renewcommand\theinnertheorem{#1}\innertheorem}
{\endinnertheorem}
\newtheorem{innercorollary}{Corollary}
\newenvironment{mycorollary}[1]
{\renewcommand\theinnercorollary{#1}\innercorollary}
{\endinnercorollary}
\newtheorem{innerproposition}{Proposition}
\newenvironment{myproposition}[1]
{\renewcommand\theinnerproposition{#1}\innerproposition}
{\endinnerproposition}
\newcommand{\minus}{\scalebox{0.75}[1.0]{$-$}}
%
%
%%% Free names, free variables 
\newcommand{\fn}[1]{\mathsf{fn}(#1)}
\newcommand{\fv}[1]{\mathsf{fv}(#1)}
%
%%% Affine Cells 
	%%%% language name 
\newcommand{\CLASS}{$\mathsf{CLASS}$} 
	%%%% CLASS without second-order and inductive/coinductive types 
\newcommand{\CLASSs}{ \CLASS{}$\setminus\exists \mu$}
	%%%% collapsing non-determinism variant 
\newcommand{\CLASSc}{$\mathsf{CLASS}_{\mathsf{c}}$} 
	%%%% Language implementaiton 
\newcommand{\CLLSj}{$\mathsf{CLLSj}$} 
%
	%%%% previous language name (exponential cells) 
\newcommand{\piSSL}{$\pi\mathsf{SSL}$} 
%
	%%%% types 
\newcommand{\aff}[1]{\wedge #1}
\newcommand{\coaff}[1]{\vee #1}
\newcommand{\cstateE}[1]{\textcolor{red}{\mathsf{S}_\circ} #1}
\newcommand{\cstateF}[1]{\textcolor{dark-blue}{\mathsf{S}_\bullet} #1}
\newcommand{\cstateFR}[1]{\textcolor{dark-blue}{\mathsf{S}} #1}
\newcommand{\cusageE}[1]{\textcolor{red}{\mathsf{U}_\circ}  #1}
\newcommand{\cusageF}[1]{\textcolor{dark-blue}{\mathsf{U}_\bullet} #1} 
\newcommand{\cusageFR}[1]{\textcolor{dark-blue}{\mathsf{U}} #1} 
\newcommand{\stateP}[2]{\textbf{S}_#1 \ #2}
\newcommand{\usageP}[2]{\textbf{U}_#1 \ #2}
\newcommand{\emptyf}{\textcolor{red}{e}} 
\newcommand{\fullf}{\textcolor{dark-blue}{f}} 
%
	%%%% constructors 
\newcommand{\affine}[2]{\textcolor{pcolor}{\mathsf{affine}} \ #1; #2} 
\newcommand{\daffine}[4]{\textcolor{pcolor}{\mathsf{affine}}_{#3, #4} \ #1; #2} 
\newcommand{\discard}[1]{\textcolor{pcolor}{\mathsf{discard}} \ #1} 
\newcommand{\use}[2]{\textcolor{pcolor}{\mathsf{use}} \ #1; #2} 
\newcommand{\cell}[3]{\textcolor{pcolor}{\mathsf{cell}} \ #1(#2.#3)} 
\newcommand{\fcell}[2]{\textcolor{pcolor}{\mathsf{cell}} \ #1(#2)} 
\newcommand{\mempty}[1]{\textcolor{pcolor}{\mathsf{empty}} \ #1} 
\newcommand{\free}[1]{\textcolor{pcolor}{\mathsf{drop}} \ #1} 
\newcommand{\user}[2]{\textcolor{pcolor}{\mathsf{use}} \ #1; #2} 
\newcommand{\uset}[0]{\textcolor{pcolor}{\mathsf{use}}} 
\newcommand{\take}[3]{\textcolor{pcolor}{\mathsf{take}} \ #1(#2);#3} 
\newcommand{\mput}[4]{\textcolor{pcolor}{\mathsf{put}} \ #1(#2.#3);#4}
\newcommand{\fput}[3]{\textcolor{pcolor}{\mathsf{put}} \ #1(#2);#3}
%
%
%%% Recursion 
	%%%% types 
\newcommand{\ind}{\textcolor{pcolor}{\mathsf{rec}}} 
\newcommand{\type}{\textcolor{pcolor}{\mathsf{type}}} 
\newcommand{\coind}{\textcolor{pcolor}{\mathsf{corec}}} 
\newcommand{\trec}[2]{\mu #1. \ #2} 
\newcommand{\tcorec}[2]{\nu #1. \ #2} 
\newcommand{\unfold}[2]{\textcolor{pcolor}{\mathsf{unfold}_\mu} \ #1; #2} 
\newcommand{\unfoldn}[2]{\textcolor{pcolor}{\mathsf{unfold}_\nu} \ #1; #2} 
\newcommand{\ncorec}[5]{\textcolor{pcolor}{\mathsf{corec}} \ #2(#1, #3) ; #5 \ [#4]} 
\newcommand{\nmuts}[5]{\textcolor{pcolor}{\mathsf{sh}}\ #1 \ #2(#3) ; #4 \ } 
\newcommand{\nmut}[5]{\textcolor{pcolor}{\mathsf{sh}}\ #1 \ #2(#3) ; #5 \ [#4]} 
\newcommand{\corec}[4]{\textcolor{pcolor}{\mathsf{corec}} \ #2(#1, #3); #4} 
\newcommand{\mloops}[3]{\textcolor{pcolor}{\mathsf{loop}} \ #2(#1); #3} 
\newcommand{\rvar}[2]{#1(#2)} 
\newcommand{\println}[2]{\textcolor{pcolor}{\mathsf{println}} \ #1; #2}
%%% Encoding 
\newcommand{\enc}[1]{\llbracket #1 \rrbracket}
%
%
%%% Refer files in implementation 
\newcommand{\impl}[1]{\emph{\text{#1}}}
%
%%% stuff for SN 
\newcommand{\Lcell}[3]{\textcolor{pcolor}{\mathsf{cell}} \ #1(#2.#3)} 
\newcommand{\Lempty}[3]{\textcolor{pcolor}{\mathsf{empty}} \ #1 (#2.#3)} 
\newcommand{\lpredE}[1]{\mathcal L \llbracket #1 \rrbracket}
\newcommand{\cred}[1]{R[#1]}
%
%%%%%%%%%%%%%%%%%%%%%%% %%%%%%%%%%%%%%%%
%%%%%%%%%%%%%%%%%%%%%%% %%%%%%%%%%%%%%%%
%%%%%%%%%%%% MY MACROS END %%%%%%%%%%%%%%%%
%%%%%%%%%%%%%%%%%%%%%%% %%%%%%%%%%%%%%%%
%%%%%%%%%%%%%%%%%%%%%%% %%%%%%%%%%%%%%%%
\maketitle
\begin{abstract}
CLASS is a proof-of-concept general purpose linear programming language, flexibly supporting realistic concurrent programming idioms, and featuring an expressive linear type system ensuring that programs (1) never misuse or leak stateful resources or memory, (2) never deadlock, and (3) always terminate. 
The design of CLASS and the strong static guarantees of its type system originates in its Linear Logic and proposition-as-types foundations. However, instead of focusing on its theoretical foundations, this paper briefly illustrates, in a tutorial form, an identifiable CLASS session-based programming style where strong correctness properties are automatically ensured by type-checking.  Our more challenging examples include concurrent thread and memory-safe mutable ADTs, lazy stream programming, and manipulation of linear digital assets as used in smart contracts.
\end{abstract}
\vspace{-10pt}
\section{Introduction}
The interpretation of linear logic as a session-typed $\pi$-calculus
\cite{DBLP:conf/concur/CairesP10,cairesmscs16,wadler2014propositions}, capturing full
session types~\cite{DBLP:conf/concur/Honda93,DBLP:conf/esop/HondaVK98,DBLP:journals/acta/GayH05}, has been intensively developed
since its proposal. %, motivating programming language design, e.g., ~\cite{DBLP:journals/lmcs/DasP22,DBLP:conf/esop/RochaC23}.
As particular relevant themes for programming language design, we may refer to
polymorphism~\cite{DBLP:conf/esop/CairesPPT13,wadler2014propositions}, 
 inductive and co-inductive types~\cite{toninho2014corecursion,DBLP:conf/esop/ToninhoY18,rocha2021propositions}, integration with higher-order functional programming and dependent types~\cite{DBLP:conf/esop/ToninhoCP13,DBLP:conf/ppdp/ToninhoCP21}, or control effects~\cite{DBLP:conf/esop/CairesP17}.
In \cite{rocha2021propositions,DBLP:conf/esop/RochaC23}, we have
introduced program 
constructs inspired by DiLL~\cite{ehrhard2018introduction}, which allow stateful shared state computation to be expressed, while keeping compatibility with the core ideology of proposition-as-types. We believe that,  pretty much like the lambda calculus is considered a canonical typed model for functional sequential computation with pure values, the linear logic typed session calculus
may be fairly considered a canonical typed model for linear stateful concurrent computation, also motivating programming language design and implementation.
Programming experience with sessions eventually led to the evolution and implementation of CLASS~\cite{rocha2021propositions,DBLP:conf/esop/RochaC23,classthesis}, a proof-of-concept general purpose linear programming language, featuring an expressive type system ensuring that programs never misuse or leak resources or memory, never deadlock, and always terminate. 
Prototype implementations have been and keep being produced, based on a fully concurrent execution model~\cite{artifactEsop23}, motivating 
our recent proposal for a fully sequential coroutine-based execution model, the Session Abstract Machine~\cite{DBLP:conf/esop/CairesT24,artifactesop24}.
% We are currently investigating efficient execution support for session programs based on the linear logic, leading to  

The strength of static guarantees often comes at the expense of a programming language's expressiveness. We believe that this is not the case for CLASS, which robustly supports many interesting ``real-world" programming idioms, involving sessions, higher-order computation, concurrency, and shared mutable state, combined elegantly under a lazy computation discipline. The aim of this paper is thus to illustrate, using examples in a tutorial format, the particular style of CLASS programming, for which all of its rather strong correctness guarantees are automatically ensured at type-checking time.

\subsection{Hello World}
Since Ritchie~\cite{DBLP:books/ph/KernighanR78}, every programming language introductory tutorial should start with the ``hello world" program. It simply prints out a greeting message and terminates. CLASS programs are collections of process definitions. On the right, we illustrate program execution at the CLASS top level REPL prompt.

\begin{center}
\begin{tabular}{ll|ll}
\begin{lstlisting}
proc main() {
  println("hello world "+(2*3));[] };;
\end{lstlisting}
&\quad&
\begin{lstlisting}
> main();;
hello world 6
\end{lstlisting}
\end{tabular}
\end{center}
\subsection{Basic Linear Session Programming}
While the core construct of functional programming is the function, in CLASS every object is a session, where a function is just a particular case of a session. Like functions, sessions are objects that may accept and return values. Unlike functions, sessions can interleave multiple steps of input and output, and offer and exercise choices. Moreover, with sessions there is no asymmetry between caller and callee,
a session describes a bidirectional interaction between two symmetric partners, described by dual types. For every session type \keyi{A} , there is its dual type $\sim$\keyi{A}. 
%Duality expresses the relationship between the two endpoints of a session. 
Typically, we use the positive form \keyi{A} for the session side that produces the object, and the negated form $\sim$\keyi{A} for the session side that consumes the object.
%
%In CLASS every object is a session and every type is a session type (even basic types, e.g., \keyc{lint}, the type of linear integers, are seen as base session types). 

We illustrate with the code for an arithmetic server object and its clients,
where process \keyi{menu} implements the server body. The protocol of the server is defined by the type \keyi{tmenu}. 

\noindent
\begin{tabular}{lllll}
\begin{lstlisting}
type tmenu {
  offer of {
    | #Dup:  recv ~lint; 
             send lint; wait 
    | #Add:  recv ~lint; recv ~lint;
             send lint; wait
}};;
\end{lstlisting}
&
&
\begin{lstlisting}
type dtmenu {
  case of {
    | #Dup:  send lint; 
             recv ~lint; close 
    | #Add:  send lint; send lint;
             recv ~lint; close
 }};;
\end{lstlisting}
\end{tabular}

It offers options \keyi{\#Dup} and \keyi{\#Add} as defined by the specific protocol. For \keyi{\#Dup},
it reads one integer, sends back its double and closes the session, while for  
\keyi{\#Add},
it reads two integers, sends back its addition and closes.
We give the  explicit definition of the dual type \keyi{dtmenu} of \keyi{tmenu}, but we could have just used
$\sim$\keyi{tmenu}.
Notice how the code for each branch option  in \keyi{menu} complies with the corresponding session type branch, and \keyc{recv} and \keyc{send} operations implement session input and output.

\noindent
\begin{tabular}{llll}
\begin{lstlisting}
proc menu(m:tmenu) {
  case m of {
    | #Dup: recv m(n);
        send m(2*n);
        wait m; []
    | #Add: recv m(n1); 
        recv m(n2);
        send m(n1+n2);
        wait m; []
  }
};;
\end{lstlisting}
&\quad
\begin{lstlisting}
proc alice0(c:~tmenu) {
  #Dup c; c <- 2; c -> m;
  println("alice got " + m);
  close c
};;

proc main0(){
  cut { menu(s)
        |s:~tmenu|
       alice0(s) }
};;
\end{lstlisting}
&\quad
\begin{lstlisting}
proc bob0(c:~tmenu){
  #Add c; 
  c <- 4; c <- 3; c -> m;
  println("bob got " + m);
  close c
};;

proc main1(){
  letc s:tmenu { menu(s) };
  bob0(s)
};;
\end{lstlisting}
\end{tabular}

Code for \keyi{main0} and \keyi{main1} exemplify the two usages of the server, one for \keyi{\#Dup} and other for \keyi{\#Add}. They essentially launch the server and compose it with client code. Sessions are composed by the \keyc{cut} construct. \keyi{main0} shows the basic CLASS cut notation from Linear Logic, but in \keyi{main1} we use the equivalent sugared \keyc{letc} notation. It defines a new interaction point (\keyi{s}) and lets the two parts of the code (server and client) interact. There is a fundamental difference between CLASS \keyc{letc} binding, and the more familiar \textbf{let} binding of functional languages, where the let expression is evaluated, bound to an identifier, and then used in the body.
In \keyc{letc} \keyi{c}:\keyi{A} \{\keyi{P}\}; \keyi{Q} the processes \keyi{P} and \keyi{Q} conceptually execute in in several interaction steps via \keyi{c}, in parallel or co-routining,  as defined by the session type \keyi{A}. Namely, in \keyi{main1()},  \keyi{s} is used at type \keyi{tmenu} in \keyi{menu(s)}, but at type \keyi{$\sim$tmenu} in \keyi{bob0(s)}. Notice in \keyi{alice0} and \keyi{bob0} the sugared notation for
send (\keyc{send} \keyi{s}\texttt{(2)} / \keyi{s} \keyc{<-} \texttt{2)} and receive (\keyc{recv} \keyi{s}(\keyi{r}) / \keyi{s} \keyc{->} \keyi{r}),
where the received object is bound to the fresh identifier \keyi{r} in the continuation. Unlike in functional languages, where function parameters always represent ``input" values, CLASS process parameters denote general session interaction points, where i/o information flow is defined at a fine grain by the session type.

\subsection{Replicated Session Programming}
Replicated (non-linear) sessions, defined in CLASS using exponential types,
may be shared like regular objects in a language like Python,
 used an unbounded (possibly zero) number of times. 
%, which allow us to promote linear code to replicated (non-linear) code. 
We illustrate here the code for a replicated arithmetic server, callable an unbounded number of times, defined by process \keyi{rserver},
and created by \keyi{rserver}(\keyi{s}) in \keyi{main}. Each time \keyi{s} is called (via a client \keyc{call} \keyi{s}(\keyi{m})), a new fresh (linear) session \keyi{m} to execute \keyi{menu(m)} at server side is spawned.
In \keyi{main}, the code for \keyi{alice} and \keyi{bob} call the same replicated server \keyi{s} (at type \keyc{?$\sim$}\keyi{tmenu}), while requesting different operations.

\begin{tabular}{lll}
\begin{lstlisting}
proc rserver(sm:!tmenu) {
    !sm(m); menu(m)
};;
\end{lstlisting}
&\quad
\begin{lstlisting}
proc bob(; rs:~tmenu) {
  call rs(b); bob0(b)
};;

proc alice(; rs:~tmenu) {
  call rs(a); alice0(a)
};;
\end{lstlisting}
&\quad
\begin{lstlisting}
proc main() {
  letc s:!tmenu {
      rserver(s)
  };
  par { alice(;s) 
        || bob(;s) }
};;
\end{lstlisting}
\end{tabular}

Notice
in process definitions the presence of parameters of implicit exponential type $\keyc{?}\keyi{A}$, declared after the ";" in the (optional) exponential context. Exponential parameters are handled non-linearly (cf. \keyi{s} in \keyi{alice(;s)} and \keyi{bob(;s)}); the only available usage for an exponential parameter \keyi{r} is \keyc{call} \keyi{r}(-).
%Execution of \keyc{call}(\keyi{c}) spans a new linear instance of the  replicated session. The sharing behaviour promoted by exponential types refer to ``pure linear objects'', which do not share state, below shared state in CLASS.
\vspace{-10pt}
\subsection{Pure Inductive Data Types and Generics}
\label{sec:flist}
Session-based programming promotes an interaction based lazy programming style 
for programs that create and manipulate linear and replicated data types.
Using recursive types and generics (universal polymorphism), we may define in CLASS a type for lists of objects of an arbitrary (session) type \keyi{A}.
\begin{center}
\begin{tabular}{lll}
\begin{lstlisting}
type rec List(A){
    choice of {
        |#Nil:  close
        |#Cons: pair A; List(A) }
};;

proc nil<A>(l:List(A)){
    #Nil l; close l
};;

proc cons<A>(a:~A, l:~List(A), nl:List(A)){
    #Cons nl; nl  <- a; fwd nl l
};;
\end{lstlisting}
&\quad
\begin{lstlisting}
proc rec concat<A>( a:~List(A),
                    b:~List(A),
                    ab:List(A)) {
  case a of {
  |#Nil: wait a; fwd b ab
  |#Cons: 
    a -> val;
    letc lx:List(A) {
      concat<A>(a,b,lx)
    };
    cons<A>(val,lx,ab)
  }
};;
\end{lstlisting}
\end{tabular}
\end{center}
The session \keyc{send}  type may be seen as a linear pair constructor (cf. in the linear logic semantics, \keyc{send} corresponds to the $ A\otimes B$ type). This explains our usage of keyword \keyc{pair} in the type of \keyi{List} as a sugared alias for \keyc{send}. Notice how negation / duality ($\sim$) in the parameters types cleanly express "input" parameters (objects consumed, rather than produced). The constructors \keyi{nil} and \keyi{cons} produce canonical linear \keyi{List(A)} values. In the \keyi{cons} case, the new list at \texttt{l} signals \keyi{\#Cons}, then exposes the element of type $A$, and then continues with the tail list \keyi{l}, as implemented by forwarding \keyi{fwd}. The concat ``function" process resembles the code we would write in a functional language, where the linear prefix list \keyi{a} is recursively 
destroyed and reconstructed. However, due to the underlying session execution model, a function like concat is executed lazily, following a concurrent or demand driven co-routing semantics.

\subsection{Lazy and Stream-based Computation}
We showcase co-recursive and affine types in 
a famous scenario of lazy computation: Turner's~\cite{turner76} filter network sieve of Eratosthenes, coded in session programming style. We first define the type \keyi{AIntStream}
of infinite streams of (non-linear) integers; 
a session of \keyc{affine} type may either by used linearly, or discarded (using \keyc{drop}). 
For convenience, we define \keyi{CIntStream}, the one step unfolding of \keyi{AIntStream}. Processes
\keyi{intsfm} and \keyi{intsfm2} generate respectively the stream of all integers from \keyi{k} and from \texttt{2}.

\noindent
\begin{tabular}{lll}
\begin{lstlisting}
type corec AIntStream {
  affine send !lint; AIntStream
};;
 
proc rec intsfm(nk: AIntStream; k:~lint)
{
  affine nk; nk <- k; intsfm(nk;k+1)
};;
\end{lstlisting}
&\quad\quad
\begin{lstlisting}
type CIntStream {
  affine send !lint; AIntStream
};;
 
proc intsfm2(n2: AIntStream)
{
  intsfm(n2;2)
};;
\end{lstlisting}
\end{tabular}

\noindent
Below we define the \keyi{filter} and \keyi{sieve} processes; \keyi{sieve} consumes the stream \keyi{sins} $n_0,n_1,\ldots$ of integers, and returns 0 for each non-prime $n_i$ and $n_p$ for each prime $n_p$ (creating a new filter for $n_p$).

\noindent
\begin{tabular}{lll}
\begin{lstlisting}
proc rec filter(fouts:AIntStream,
                fins:~CIntStream; n:~lint)
{
    fins -> v;  
    if (v mod n == 0) then {
        affine fouts;
        fouts <- 0; 
        filter(fouts,fins;n) 
    } else {
       affine fouts;
       fouts <- v;
       filter(fouts, fins; n) 
    }
};;
\end{lstlisting}
&\quad\quad
\begin{lstlisting}
proc rec sieve(souts:AIntStream,
               sins:~CIntStream) {
  sins -> p;
  affine souts;
  souts <- p;
  if (p == 0) {
    sieve(souts,sins)
  } else {
    letc outp:AIntStream  {
      filter(outp,sins;p)
    };
    sieve(souts,outp)
  }
};;
\end{lstlisting}
\end{tabular}

We now present sample driver code, in this case 
\keyi{main_sa(;n)}
prints the primes up to \keyi{n}.

\noindent
\begin{tabular}{lll}
\begin{lstlisting}
proc primesN(lp:AIntStream)
{
  letc ln:AIntStream { intsfm2(ln) };
  sieve(lp,ln)
};;

proc main_sa(;n:~lint)
{
  letc lp:AIntStream 
    { primesN(lp) };
  print2k(lp;n)
};;
\end{lstlisting}
&\quad\quad
\begin{lstlisting}
proc gen_rec print2k(il:~AIntStream;
                       k :~lint)
{
  if(k==1) then { println(""); drop il }
  else {
    il -> n;
    if (n==0) { 
      print2k(il;k-1)
    } else { print(n+" "); print2k(il;k-1)
    }
  }
};;
\end{lstlisting}
\end{tabular}

While \keyi{filter} and \keyi{sieve} are strictly inductively defined processes (termination-safe by typing), we illustrate in \keyi{printup2k} the use of general recursion in a ``while'' type iteration. CLASS allows \keyt{gen_rec} as an unsafe escape mechanism, useful to write non-terminating or general recursive programs. Using \keyt{gen_rec} we developed a library of well-typed termination-safe iterators and generators (cf. Python's range()), which we prefer not to use here for clarity in showcasing pure CLASS code.
\subsection{Shared Mutable State}
A key feature of CLASS is safe manipulation of shared linear state~\cite{rocha2021propositions,classthesis,DBLP:conf/esop/RochaC23}, supporting behaviourally typed reference cells. These turn out to match a typed version of Concurrent Haskell's MVars~\cite{jones1996concurrent}. The type \keyc{state} \keyi{A} denotes the type of cells holding objects of type \keyi{A}, with dual type $\keyc{usage} {\sim} \keyi{A}$. Cells are manipulated with \keyc{take}, \keyc{put} and \keyc{drop} operations: \keyc{take} moves the cell-stored object to the reader (linear move semantics, emptying the cell), \keyc{put} moves an object from the caller into the cell (linear move semantics, filling the cell), and \keyc{drop} releases cell reference usage (cf. session \keyc{close}), causing the cell to be deallocated when no more references to it exist (our implementation uses reference counting). Types ensure any cell alias is used according to a linear protocol suggested by the regex \texttt{(take;put)*;drop}.

\noindent
\begin{tabular}{lll}
\begin{lstlisting}
proc main0m() {
  letc m:state Int {
    cell m(42)
  };
  take m(x);
  println(x);
  put m(x);
  drop m
}
;;
\end{lstlisting}
&\quad\quad
\begin{lstlisting}
proc main1m() {
  letc m:state !lint {
    cell m(2)
  };
  share m {
    take m(x); put m(x+1); println(x); drop m  
    ||
    take m(x); put m(x-1); println(x); drop m  
  }
};;
\end{lstlisting}
\end{tabular}

The code \keyi{main0m} exemplifies a simple linear usage of a reference cell. The code for \keyi{main1m} illustrates sharing of reference cells, made here explicit by the \keyc{share} construct. The reference cell \keyi{m} is shared by two independent threads, that concurrently interleave cell operations non-deterministically, where shared \keyc{take}s require acquiring a cell mutex, to be eventually released by \keyi{put}. Hence, \texttt{main1m} prints \texttt{2}  followed by either \texttt{3} or \texttt{1}.
Type-checking of  \keyc{share} (resp. \keyc{cut}) ensures that each of the two
independent threads involved can at most share one reference cell (resp. session). This discipline is crucial to ensure deadlock absence and does not hinder expressiveness~\cite{DBLP:conf/esop/RochaC23}. When all threads using some shared cell drop it, the
cell is deallocated (as well as its disposable content, required to be either of \keyc{affine} or \keyc{state} type). 
Sharing annotations do not have any special operational meaning, and may be inferred by the CLASS interpreter elaboration phase, but here we manifest in the code  all occurrences of sharing via \keyc{share}, for clarity. 
Notice that \keyc{share} is not a static scoping construct; cell references may be freely passed around, like any all other session objects, as we exemplify in the code snippets below.
\begin{center}
\begin{tabular}{lll}
\begin{lstlisting}
type Mint { state lint };;
type HO { send Mint; wait };;

proc sender(s:HO) {
  letc m:Mint { cell m(2) };
  share m {
    s <- m; wait s; ()
    || 
    take m(v); put  m(v+1); drop m
  }
};; 
\end{lstlisting}
&\quad
\begin{lstlisting}
proc receiver(s:~HO) {
  s -> c;
  take c(v);
  println (v);
  put c(0);
  drop c;
  close s
};;
\end{lstlisting}
&\quad
\begin{lstlisting}
proc pass() {
  letc s:HO {
    sender(s)
  };
  receiver(s)
};;
\end{lstlisting}
\end{tabular}
\end{center}
Program \keyi{pass} composes \keyi{sender} and \keyi{receiver} via a session of type \keyi{HO} (standing for HandOver). The \keyi{sender} allocates a fresh cell \keyi{m} and sends (an alias of) \keyi{m} to \keyi{receiver}, while locally 
 increments it via the retained alias \keyi{m} before dropping it. Concurrently, the \keyi{receiver} reads the cell contents, sets it to $0$, and drops it. When the program terminates, neither the cell has been leaked, nor any reference to it became dangling, as ensured by construction of the typed CLASS code.
%Our following examples are more elaborated, and illustrate CLASS programming of some very realistic concurrency abstractions.
\subsection{A Concurrent Barrier for $N$ Threads}
We implement in CLASS a generic barrier abstraction (see e.g., Rust 
\textsf{std::sync::Barrier}~\cite{rust-lang}).
The data representation of the barrier object is a (shared) memory cell storing a 
pair (type \keyi{BState}) 
of an integer counting the number of threads yet to synchronise
and a list of waiting thread continuations (we reuse \keyi{List(A)} from Section~\ref{sec:flist}). We model a thread continuation as a session object of type \keyi{Cont}, simply waiting for a \keyc{close} signal to run.
%A barrier (type \keyi{Barrier}) is represented by a (shared) memory cell storing a \keyi{BState}.
%
Process \keyi{barrier} creates a new barrier \keyi{b} for \keyi{nt} threads,
 auxiliary process \keyi{init} sets up the initial state, with the
 counter set to \keyi{nt} and an empty waiting list.
The process \keyi{awakeall} is called 
when all threads reach the barrier; it traverses the waiting list \keyi{ws} and launches
(via \keyc{close} \keyi{w}) each pending continuation.

\begin{tabular}{lll}
\begin{lstlisting}
type Cont { affine wait };;
type BState
  { pair !lint; affine List(Cont) };;
  
type Barrier { state BState };;

proc init(rep: BState;n:~lint){
    rep <- n;
    affine rep; 
    nil<Cont>(rep)
};;
\end{lstlisting}
&\quad
\begin{lstlisting}
proc barrier(b:Barrier;nt:~lint){
    cell b(r. affine r; init(r;nt))
};;

proc rec awakeAll(ws:~List(Cont)){
  case ws of {
    | #Nil : wait ws;()
    | #Cons: recv ws(w);
             close w; 
             awakeAll(ws)
    }};;
\end{lstlisting}
\end{tabular}

The core of the barrier code is process \keyi{bwait}.
Any thread about to register in the barrier
calls \keyi{bwait}, 
passing its owned shared reference to the barrier object cell at \keyi{b},
and its own continuation at \keyi{cont}. 

\noindent
\begin{tabular}{lll}
\begin{lstlisting}
proc bwait(b:~Barrier, cont:~Cont) {
  take b(ws);
  ws -> n;
  if n==1 then {
    par { awakeAll(ws)
          ||
        put b(nw. affine nw; init(nw;0));
        close cont; release b
    }
  } else {
   letc nw: affine BState {
     affine nw; nw <- n-1;  
     affine nw; cons<Cont>(cont,ws,nw) };
     put b(nw); drop b
  } 
};;
\end{lstlisting}
&
\begin{lstlisting}
proc thread(b:~Barrier; i:~lint) {
  println("thread " + i + " started.");
  sleep 99; // work before barrier
  letc cont: ~affine wait {	    
    println("thread " + i + " on wait");
    bwait(b,cont) // call barrier wait
  };
  affine cont; 
  wait cont; // wait here
  println("thread " + i + " wake up.");
  sleep 101; // work after barrier
  println("thread " + i + " terminates.");
  []
}
;;
\end{lstlisting}
\end{tabular}

Each time \keyi{bwait} is called, it takes  the barrier state (acquiring the mutex).
If the calling thread is the last one \texttt{(\keyi{n}==1)}, it concurrently awakes all waiting threads, while, in parallel, launches the calling thread and drops its ownership of the barrier (for simplicity, we assume the barrier to be single use).

Otherwise, it adds the continuation to the waiting queue (using \keyi{cons}), decrements the count of threads that did not reach the barrier yet, and updates the state accordingly using \keyc{put} \keyi{b}(\keyi{nw}), which releases the barrier mutex.
We simulate the job of each thread \keyi{i} with process \keyi{thread},
which does some prior work (\keyc{sleep} \texttt{99}), and calls \keyi{bwait} with
a continuation to do some after work (\keyc{sleep} \texttt{101}).
Notice that some types are declared \keyc{affine}; recall that a session of \keyc{affine} type may either by used linearly, or discarded, and that values storable in cells are required to be of \keyc{affine} or \keyc{state} type. 
We conclude our concurrent barrier example with some client code.  

\noindent
\begin{tabular}{lll}
\begin{lstlisting}
proc gen_rec spawnall(b:~Barrier; i:~lint, n:~lint) {
  if n == 0 then { drop b }
  else
    { share b { thread(b;i) || spawnall(b;i+1,n-1) }
  }
};;
\end{lstlisting}
&
&
\begin{lstlisting}
proc mainb(;nt:~lint) {
    letc c: Barrier {
    	barrier(c;nt)
    };
    spawnall(c;0,nt)
};;
\end{lstlisting}
\end{tabular}

The main program \keyi{mainb} creates a new barrier for \keyi{nt} threads and concurrently launches the code for each one, calling \keyi{spawnall}.  
All the core code for the barrier above is deemed thread-safe by typing: no deadlocks, livelocks, or memory leaks may arise.

\subsection{(Lazy) Mutable Data Structures}
\label{sec:llist}
We show an implementation for lists of linked memory cells, using a tail sentinel node.
Each list element is a memory cell with content as specified by type 
\keyi{ANode(A)} (CLASS annotates the argument type of a \keyc{state} type constructor automatically as \keyc{affine} except already explicitly typed \keyc{affine} or \keyc{state}). Notice that definition of types
\keyi{LList(A)} and \keyi{Node(A)} is mutually recursive.
\begin{center}
\begin{tabular}{lll}
\begin{lstlisting}
type rec LList(A) { state Node(A) }
  and Node(A) { choice of {
                  | #Nil : close
                  | #Next : pair affine A; LList(A)
                 }
};;

proc cons<A>(a:~affine A, t:~LList(A), l: ANode(A)){
  affine l; #Next l; l <- a; fwd l t
};;
\end{lstlisting}
&\quad
\begin{lstlisting}
type ANode(A) {
    affine Node(A)
};;


proc nil<A>(l: ANode(A)) {
    affine l;
    #Nil l;
    close l
};;
\end{lstlisting}
\end{tabular}
\end{center}
The code for \keyi{cons} builds at \keyi{l} a new \keyi{ANode}, pairing a
 value \keyi{a} of type \keyi{A} with a reference  to the list tail
 \keyi{l}.

\begin{center}
\begin{tabular}{lll}
\begin{lstlisting}
proc rec concat<A>(a:~LList(A), b:~LList(A), ab: LList(A)){
    take a(node);
    case node of {
        | #Nil: put a(n.nil<A>(n)); fwd b ab
        | #Next: node -> val;
                 letc nodeb:LList(A) { concat<A>(node,b,nodeb) };
                 put a(node. cons<A>(val,nodeb,node)); fwd a ab
    }
};;
\end{lstlisting}
&\quad
\begin{lstlisting}
\end{lstlisting}
\end{tabular}
\end{center}
This \keyi{concat} ``function" process resembles the code we would write in 
an imperative C-like language. Due to the take / put move semantics, the list is traversed recursively 
by following the references by taking cells on down calls and putting back on returns, until the last node is reached, causing the sentinel to be updated in place.
However, due to the lazy session execution model, which uniformly follows a demand driven co-routing semantics,  the concatenation of two lists is done in O(1) time, 
with reconstruction of the possibly shared imperative structure 
amortised in future transversals. We challenge the reader to wonder what   happens when a well typed shared \keyi{LList(A)} is concurrently concatenated with other lists.
\subsection{An Abstract Data Type of Digital Assets}
In this example, we illustrate state encapsulation using behavioral interfaces to code 
a linear abstract data type representing a leak-free wallet of digital tokens, as used e.g., in a blockchain app. Types \keyi{IWallet(X)} and \keyi{IWallet(X)} are mutually defined.
We pick \keyi{List(X)} as representation type (for generic token type \keyi{X}), and define the external interface by the co-recursive type \keyi{IWallet(X)}.
\begin{center}
\begin{tabular}{lll}
\begin{lstlisting}
type corec IWallet(X) {
    offer of {
        |#Count: send !lint; IWallet(X)
        |#Add: recv ~X; IWallet(X)
        |#Get: Ans(X)
    } 
\end{lstlisting}
&\quad
\begin{lstlisting}
} and Ans(X) {
    choice of {
        |#Some: send X; IWallet(X)
        |#None: close
    }
};;
\end{lstlisting}
\end{tabular}
\end{center}
Notice that ``method'' \keyc{\#}\keyi{Count} returns the number of stored tokens, 
\keyc{\#}\keyi{Add} adds a new token, and \keyc{\#}\keyi{Get} extracts a token,
if the wallet is empty, \keyc{\#}\keyi{Get} returns \keyc{\#}\keyi{None} and the wallet terminates (is disposed). 
\begin{center}
\begin{tabular}{lll}
\begin{lstlisting}
proc rec tokens_imp<A>(tm:IWallet(A), 
                          st:~List(A)) {
  case tm of {
  |#Count: 
    letc rc: { len<A>(st,rc) };
    rc -> ns; tm <- rc;
    tokens_imp<A>(tm,ns)    
  |#Add:
    tm -> val; 
    letc ns: { cons<A>(val,st,ns) };
    tokens_imp<A>(tm,ns)    
   |#Get: 
     case st of {
     |#Nil: wait st; #None tm; close tm
     |#Cons: st -> val; #Some tm; tm <- val; 
        tokens_imp<A>(tm,st)       
     }
  }
};;
\end{lstlisting}
&\quad
\begin{lstlisting}
proc rec len<A>(a:~List(A),
                ao:pair List(A);!lint)
{
...
};;

proc newTokens(tm:IWallet(lstring)) {
    letc s: { nil<lstring>(s) };
    tokens_imp<lstring>(tm,s)
};;

proc test(tk:IWallet(lstring)) {
    letc t: { newTokens(t) };
    #Add t; t <- "NFT@A36D54F89606A";
    #Count t; 
    t -> n;
    println ("balance = "+n);
    fwd t tk
};;
\end{lstlisting}
\end{tabular}
\end{center}
The wallet behaviour is implemented by the recursive process \keyi{tokens_imp} at
session \keyi{tm} ,
at each step it branches on the selected ``method'', executes the operations and recurses updating the state passed in \keyi{st}.
We leave as exercise to the reader the definition of the code for procedure \keyi{len}: given the linear list \keyi{a}, it should return at \keyi{ao} the same list and its length
(as a \keyc{!lint}). Usage by client code of an object of type \keyi{IWallet} is only possible via its interface type, as illustrated in \keyi{test}. The depicted code ensures that the representation state is never tampered with, and tokens never duplicated, erased or double spent, by linearity and parametricity, since no such capabilities are exported by the ADT.

CLASS type system also supports existential types, which may be used to express more flexible modes of information hiding, but which we are unable to cover in the present brief overview. 
\subsection{A Mutable Shared Queue}
We now address a more challenging concurrent programming exercise, 
sometimes used as a benchmark for formal verification techniques. We code in CLASS 
a shared concurrent LIFO queue offering O(1) enqueue and dequeue 
operations using the mutable linked list data structure of Section~\ref{sec:llist}. For simplicity, we assume that the queue stores \keyi{Aint} typed values.
It implements two separate usage interfaces:
one of type \keyi{EnqI} (for enqueing) and other of type \keyi{DeqI} (for dequeing),
encapsulating shared state and code. Both (corecursively typed) views allow clients to call the operations until droping (\keyi{\#Drop}) the reference.
\begin{center}
\begin{tabular}{lll}
\begin{lstlisting}
type corec EnqI {
  offer of {
    | #Enq: recv ~affine lint; EnqI
    | #Drop: wait
  }};;
\end{lstlisting}
&\quad
\begin{lstlisting}
type corec DeqI {
  offer of {
    | #Deq: pair Opt(Aint); DeqI
    | #Drop: wait
  }};;
\end{lstlisting}
\end{tabular}
\end{center}
In the case for \keyi{\#Deq}, we return an option type: when the queue is empty, the dequeue operation will return \keyi{\#None} (see the the short definition of the \keyi{Opt(A)} type below). The queue representation will maintain two cells of type \keyi{Ptr}, one for the head of the list, for dequeing, and other for the last sentinel (empty) list element (for enqueuing). Recall that the list will always contains a ``dummy" 
sentinel element at the tail. Each list node will be an object of type 
\keyc{state} \keyi{Node(lint)}.
\begin{center}
\begin{tabular}{lllll}
\begin{lstlisting}
type Aint
  { affine lint };;

type Opt(A) {
  affine choice of {
    | #None : close
    | #Some : A
  }
};;
\end{lstlisting}
&
&
\begin{lstlisting}
proc None(o: Opt(Aint)) {
  affine o;
  #None o;close o
};;

proc Some(val:~Aint,
          o:Opt(Aint)) {
  affine o;
  #Some o; fwd val o
};;
\end{lstlisting}
&
&
\begin{lstlisting}
type Ptr
  { state
    state Node(lint) };;

proc free(p:~statel
            Node(lint))
{
  put p(c. nil<lint>(c));
  drop p
};;

\end{lstlisting}
\end{tabular}
\end{center}
The code for \keyi{deq} accesses the contents \keyi{lh} of the cell
\keyi{hp} at the head,
and inspects it. If set to \keyi{\#Nil}, it is the
sentinel node (empty queue); the content of \keyi{hp} is reset, and \keyi{\#None}  returned
(the sugared closure notation \keyi{rv} \keyc{<-}
\{\keyi{r}.\keyi{Node}(\keyi{r})\}
represents \keyc{send} \keyi{rv} \textsf{\{\keyi{r}.\keyi{Node}(\keyi{r})\};}).
The code for \keyi{enq} allocates a fresh sentinel 
node \keyi{nn}. It then stores the value to enqueue \keyi{v}
and the reference to \keyi{nn} at the sentinel node \keyi{sn} (using \keyi{cons}, and an update in place).
It also stores the reference \keyi{nn} as the new sentinel node in \keyi{tl}. At the end, \keyi{nn} is shared 
by the \keyi{\#Next} field of the last queue node 
and by the tail cell \keyi{tl}. Notice the crucial use of \keyc{share}, ensuring safe dynamic sharing of state between the interfering access paths from queue head and tail. 
\begin{center}
\begin{tabular}{lll}
\begin{lstlisting}
proc deq(hd:~Ptr, rv:pair Opt(Aint);Ptr) {
  take hd(hp);
  take hp(lh);
  case lh of {
    |#Nil : wait lh;
          put hp(c. nil<lint>(c));
          put hd(hp);
          rv <- { r. None(r) }; fwd hd rv
    |#Next : 
          recv lh(val);
          rv <- { r. Some(val,r) };
          put hd(lh); free(hp); fwd hd rv
    }
};;
\end{lstlisting}
&\quad
\begin{lstlisting}
proc enq(tl:~Ptr,
         v:~affine lint, tlo:Ptr) {
  letc nn:LList(lint) {
    cell nn (c. nil<lint>(c))
  };
  take tl(sn);
  share nn {
    take sn(lp);
    put sn(c. cons<lint>(v,nn,c));
    drop sn; discard lp
    ||
    put tl(nn); fwd tl tlo
  }
};;
\end{lstlisting}
\end{tabular}
\end{center}
The queue constructor \keyi{lqueue}
 builds the initial structure of a (empty) queue.
It allocates and initialises the (empty) sentinel list node \keyi{sn}, and stores a shared
reference to it in the head \keyi{hd} and tail \keyi{hd} cells. The queue interfaces are then offered
by the \keyi{deqop} and \keyi{enqop} processes, which ``bind" the \keyi{DeqI} and \keyi{EnqI} session protocols to the actual implementation of the queue operations \keyi{deq} and \keyi{enc}. 
\begin{center}
\begin{tabular}{lll}
\begin{lstlisting}
proc lqueue(ienq:EnqI,ideq:DeqI) {
  letc sn: LList(lint) {
    cell sn (c.nil<lint>(c)) };
  share sn {
        letc hd:Ptr { cell hd (sn) }; deqop(ideq,hd)
        ||
        letc tl:Ptr { cell tl (sn) }; enqop(ienq,tl)
    }
};;
\end{lstlisting}
\end{tabular}
\end{center}

\begin{center}
\begin{tabular}{lll}
\begin{lstlisting}
proc rec deqop(deci:DeqI,tl:~Ptr) {
  case deci of {
    |#Deq:  letc tnext:pair Opt(Aint); Ptr {
              deq(tl,tnext)
            };
            recv tnext (val);
            deci <- val;
            deqop(deci,tnext)
     |#End: wait deci; 
            drop tl
    }
};;
\end{lstlisting}
&\quad
\begin{lstlisting}
proc rec enqop(enqi:EnqI,tl:~Ptr) {
  case enqi of {
    |#Enq:  recv enqi(item); 
            letc tnext:Ptr {
              enq(tl,item,tnext)
            };         
            enqop(enqi,tnext)
    |#End:  wait enqi; 
            drop tl
    }
    
};;
\end{lstlisting}
\end{tabular}
\end{center}
This last example highlights some key insights about how CLASS type system compositionally and implicitly captures non-interference and acyclicity in 
programs' data and control structures.

The code for all examples in this paper
may be found at the web site~\cite{classsiteist}.
\hide{Moreover, the fact that the clean code 
in our examples type-checks implies complete thread and memory safety of our examples implementation, showcasing the singular expressiveness of CLASS linear typing discipline.}

\hide{
\noindent
\begin{tabular}{lll}
\begin{lstlisting}
proc gen_rec deq_c(cl:~DeqI;N:~lint) {
  if N==0 then { #End cl; close cl }
    else {
    #Deq cl; 
     recv cl(ans);
      case ans of {
        | #None:
             println("NONE"); wait ans;
             deq_c(cl;N-1)
        | #Some:
             println("deq "+ans);
             deq_c(cl;N-1)
        }
    }
};;\end{lstlisting}
&\quad
\begin{lstlisting}
proc gen_rec enq_c(cl:~EnqI;N:~lint) {
  if N==0 then { #End cl; close cl }
  else {
    #Enq cl;
    cl <- N;
    println("enq "+N);
    enq_c(cl;N-1)
  }
};;

proc mainq() {
    letc cl1:~EnqI { enq_c(cl1;2048)};
    letc cl2:~DeqI { deq_c(cl2;2048)};
    lqueue(cl1,cl2)
};;
\end{lstlisting}
\end{tabular}
}
\vspace{-10pt}
\section{Concluding Remarks}
We have presented a brief tutorial on the key
design principles and features of CLASS language, illustrating its expressiveness in realistic
concurrent session-based and shared-state programs. Details more technically focused on the development and foundations of CLASS may be found in~\cite{rocha2021propositions,DBLP:conf/esop/RochaC23,classthesis,classsiteist}.

Besides the examples in this paper, many more functional, imperative and concurrent  shared state CLASS code has been developed, and automatically type-checked for the strong safety and liveness properties ensured by its linear type system.
Our experience has shown that CLASS type system, despite its expressive power, is flexible enough to deal, without resorting to unsafe features, with complex resource acquisition protocols, such as the dynamic dining philosophers, or the complete ``low-level''  implementation of Hoare-style monitors with condition variables.

Linear types are becoming more and more relevant in computing practice, as witnessed by the widespread adoption of programming languages such as Rust~\cite{rust-lang}, for general systems programming, Move~\cite{DBLP:conf/cav/ZhongCQGBPZBD20}, for blockchain smart contracts, Linear Haskell~\cite{DBLP:journals/pacmpl/BernardyBNJS18}, and other ~\cite{DBLP:journals/pacmpl/JacobsB23,DBLP:journals/corr/abs-1904-01284,DBLP:journals/lmcs/DasP22,DBLP:conf/ecoop/ChenBT22,DBLP:conf/popl/LangeNTY17}. We expect that some of the ideas introduced in CLASS to be of quite wide application.
More information about CLASS, its foundations, current versions of various implementations, and coding examples has been maintained in our evolving
web site~\cite{classsiteist}. Ongoing work on a CLASS compiler for a LLVM/CLANG backend is expected to support a fair assessment of CLASS' performance~\cite{antunes2025}. 

I would like to thank Bernardo Toninho, Frank Pfenning, Pedro Rocha, Ricardo Antunes, Vasco T. Vasconcelos, Philip Wadler, Sam Lindley, Jorge A. Perez, Nobuko Yoshida, Stephanie Balzer, 
and Peter Thiemann for many related discussions, the anonymous referees for very useful comments, and project BIG H22020 Grant ID 952226 for supporting this research.

\bibliographystyle{eptcs}
\bibliography{bibliography2}
\end{document}